\documentclass[prb,aps,twocolumn,superscriptaddress,showpacs,preprintnumbers,amsmath,amssymb]{revtex4}
\usepackage{color}
\usepackage{graphicx}
\usepackage{ulem}
\usepackage{array}
\usepackage{url}
\usepackage{enumitem}
\usepackage[colorlinks=true,urlcolor={blue},citecolor={blue}]{hyperref}
\usepackage{bbold}

\def\textbf#1{\boldsymbol{#1}}

\newcommand{\upperRomannumeral}[1]{\uppercase\expandafter{\romannumeral#1}}
\newcommand{\lowerromannumeral}[1]{\romannumeral#1\relax}

\begin{document}

\title{Superconductivity-induced Phonon Renormalization on NaFe$_{1-x}$Co$_{x}$As}

\author{Y.J.~Um}
\affiliation{Max-Planck-Institut~f\"{u}r~Festk\"{o}rperforschung,
Heisenbergstr.~1, D-70569 Stuttgart, Germany}

\author{Yunkyu~Bang\footnote{Corresponding author: ykbang@chonnam.ac.kr}}
\affiliation{Department of Physics, Chonnam National University,
Kwangju 500-757, Korea}

\author{B.H.~Min}
\affiliation{Department of Emerging Materials Science, Daegu Gyeongbuk Institute of Science \& Technology
, Daegu 711-873, Republic of Korea}

\author{Y.S.~Kwon\footnote{Corresponding author: yskwon@skku.ac.kr}}
\affiliation{Department of Emerging Materials Science, Daegu Gyeongbuk Institute of Science \& Technology
, Daegu 711-873, Republic of Korea}

\author{M.~Le Tacon\footnote{Corresponding author:m.letacon@fkf.mpg.de}}
\affiliation{Max-Planck-Institut~f\"{u}r~Festk\"{o}rperforschung,
Heisenbergstr.~1, D-70569 Stuttgart, Germany}

\date{\today}

\begin{abstract}
We report a study of the lattice dynamics in superconducting NaFeAs ($T_c$ = 8 K) and doped NaFe$_{0.97}$Co$_{0.03}$As ($T_c$ = 20 K) using Raman light scattering. Five of the six phonon modes expected from group theory are observed. In contrast with results obtained on iso-structural and iso-electronic LiFeAs, anomalous broadening of E$_{g}$(As) and A$_{1g}$(Na) modes upon cooling is observed in both samples. In addition, in the Co-doped sample, a superconductivity-induced renormalization of the frequency and linewidth of the B$_{1g}$(Fe) vibration is observed.
This renormalization can not be understood within a single band and simple multi-band approaches. A theoretical model that includes the effects of SDW correlations along with sign-changing $s$-wave pairing state and interband scattering has been developed to explain the observed behavior of the B$_{1g}$(Fe) mode.

\end{abstract}


\maketitle

\section{Introduction}

The discovery of superconductivity in LiFeAs~\cite{Tapp_PRB2008, Wang_SSC2008} has triggered several debates in the field of iron-based superconductors research. 
Most of the iron-based superconductors known so far share similar properties, namely (\lowerromannumeral{1}) a magnetic transition accompanying a structural transition in stoichiometric parent compounds, (\lowerromannumeral{2}) a strongly nested-Fermi surface that induces a SDW instability, and (\lowerromannumeral{3}) the appearance of superconductivity upon chemical doping or application of external pressure.
Even though LiFeAs has a crystal and electronic structures very similar to those of other families of iron-based superconductors~\cite{Tapp_PRB2008, Wang_SSC2008,Chu_PC2009,Li_PRB2009}, neither structural, nor magnetic phase transitions have been reported.  Furthermore it becomes superconducting at $T_c$ = 18 K at ambient pressure and without any doping.
Superconductivity with $T_c \sim$ 10 K has also been found in isostructural and isoelectronic NaFeAs~\cite{Chu_PC2009,Parker_CC2009}. However, contrary to LiFeAs,
a structural - tetragonal to orthorhombic - phase transition occurs at $T_{S} \sim$ 52 K~\cite{Parker_PRL2010,Chen_PRL2009} in NaFeAs, which further orders magnetically below $T_{SDW} \sim$ 41 K. Another difference between the two compounds is that unlike LiFeAs,  where superconductivity is suppressed upon chemical doping~\cite{Pitcher_JACS2010}, the superconducting transition temperature in NaFeAs is enhanced when substituting Ni or Co to Fe~\cite{Parker_PRL2010,Tanatar_PRB2012, Spyrison_PRB2012}. With such properties, NaFeAs is a promising candidate to bridge the gap between LiFeAs and other families of iron-based superconductors.

There is now a consensus that Fe-based superconductors are non-conventional superconductors, in which the electron-phonon interaction plays only a minor role~\cite{Boeri_PRL2008}. Several experimental studies of the phonon spectra in these compounds have however shown significant deviations from the results of standard density function theory (DFT) calculations~\cite{LeTacon_PRB2008, Reznik_PRB2009, LeTacon_PRB2009, Fukuda_JPSJ2008, Fukuda_PRB2011, Mittal_PRL2009, Hahn_PRB2009, Delaire_PRB2010, LeTacon_JPCS2011}, which can only be explained including explicitly magnetism (even in the absence of long-range order) in the lattice dynamical calculation. In relation to this, it has been shown that the magnetic interaction could enhance the electron-phonon coupling~\cite{Yildirim_PhysicaC2009, Boeri_PRB2010, Li_JAP2012, Garcia_PRB2013, Yndurain_EPL2011}. Experimental manifestations of the electron-phonon coupling in Fe-based compounds remain however marginal, and with only a few exceptions~\cite{Choi_JPCM2010,Litvinchuk_PRB2011}
, no phonon anomalies are seen across the superconducting transition with infra-red~\cite{Hu_PhysicaC2009, Dai_ChinesePhysicsB2012} or Raman~\cite{Litvinchuk_PRB2008,Rahlenbeck_PRB2009,Um_PRB2012_FTS,Um_PRB2012_LFA,Chauviere_PRB2009} spectroscopies (whereas rather large anomalies are seen across the magneto-structural transitions~\cite{Choi_PRB2008,Gnezdilov_PRB2011,Akrap_PRB2009,Chauviere_PRB2009,Rahlenbeck_PRB2009,Charnukha_PRB2013}).

In this paper, we focus on the phonons at the zone center, and report an experimental study of the lattice dynamics in NaFe$_{1-x}$Co$_{x}$As (x = 0, 0.03) single crystals using Raman scattering.
We have observed five of the six phonon modes expected from group theory and found several unusual features. This includes the anomalous broadening of E$_{g}$(As) and A$_{1g}$(Na) modes at low temperature, and a superconductivity-induced renormalization of the B$_{1g}$(Fe) mode in the Co-doped compound that can not be understood within a single band and simple multiband approaches.
We show that it can be accounted for in a theoretical framework that explicitly includes a sign-changing order parameter as well as interband scattering in presence of SDW correlations.

\section{Experimental Details}

Single crystals of parent NaFeAs ($T_c$ = 8 K) and optimally doped NaFe$_{0.97}$Co$_{0.03}$ ($T_c$ = 20 K) were grown by the Bridgman method~\cite{Song_APL2010,Park_PRB2012}. Na lump and Fe$_{1-x}$Co$_x$As precursor were mixed with appropriate stoichiometric ratio. The Fe$_{1-x}$Co$_x$As precursor was synthesized by reacting Fe and As pieces at 1050~$^\circ$C for five days. Then, in a second step, all the required ingredients were weighed and mixed in the glove box filled with He gas and put into a molybdenum crucible. The crucible was sealed through arc welding under Ar-gas atmosphere. Large and shiny crystals of 5 $\times$ 5 $\times$ 5 mm$^3$ or more were grown at 1450~$^\circ$C. The $c$-axis lattice parameter was measured using x-ray diffraction (not shown here) and estimated to 7.062\AA~and 7.045\AA~for NaFeAs and NaFe$_{0.97}$Co$_{0.03}$, respectively, in good agreement with a previous report~\cite{Parker_PRL2010}.

Prior to the Raman measurements, the air sensitive NaFe$_{1-x}$Co$_x$As single crystals were cleaved and mounted on the cold finger of a helium-flow cryostat in a glove box under Ar atmosphere.
Raman spectra have been measured in backscattering geometry and recorded with a JobinYvon LabRam 1800 single grating spectrometer equipped with a razor-edge filter and a Peltier-cooled CCD camera. We used the $\lambda$ = 632.817 nm line of a He$^{+}$/Ne$^{+}$ mixed gas laser for excitation. The laser beam was linearly polarized and focused through a 50$\times$ microscope objective to a $\sim$ 5 $\mu$m diameter spot with less than 1 mW power on the sample surface to avoid laser-induced heating. In order to determine the precise frequency of phonons for each temperature, Neon emission lines were recorded between the measurements. For the data analysis, all phonon peaks were fitted by Lorentzian profiles, convoluted with the spectrometer resolution function (a gaussian line of 2 cm$^{-1}$ full width at half maximum (FWHM)).


\section{Experimental Results}
\subsection{Mode Assignment in parent NaFeAs}

\begin{figure}
\includegraphics[width=1\linewidth]{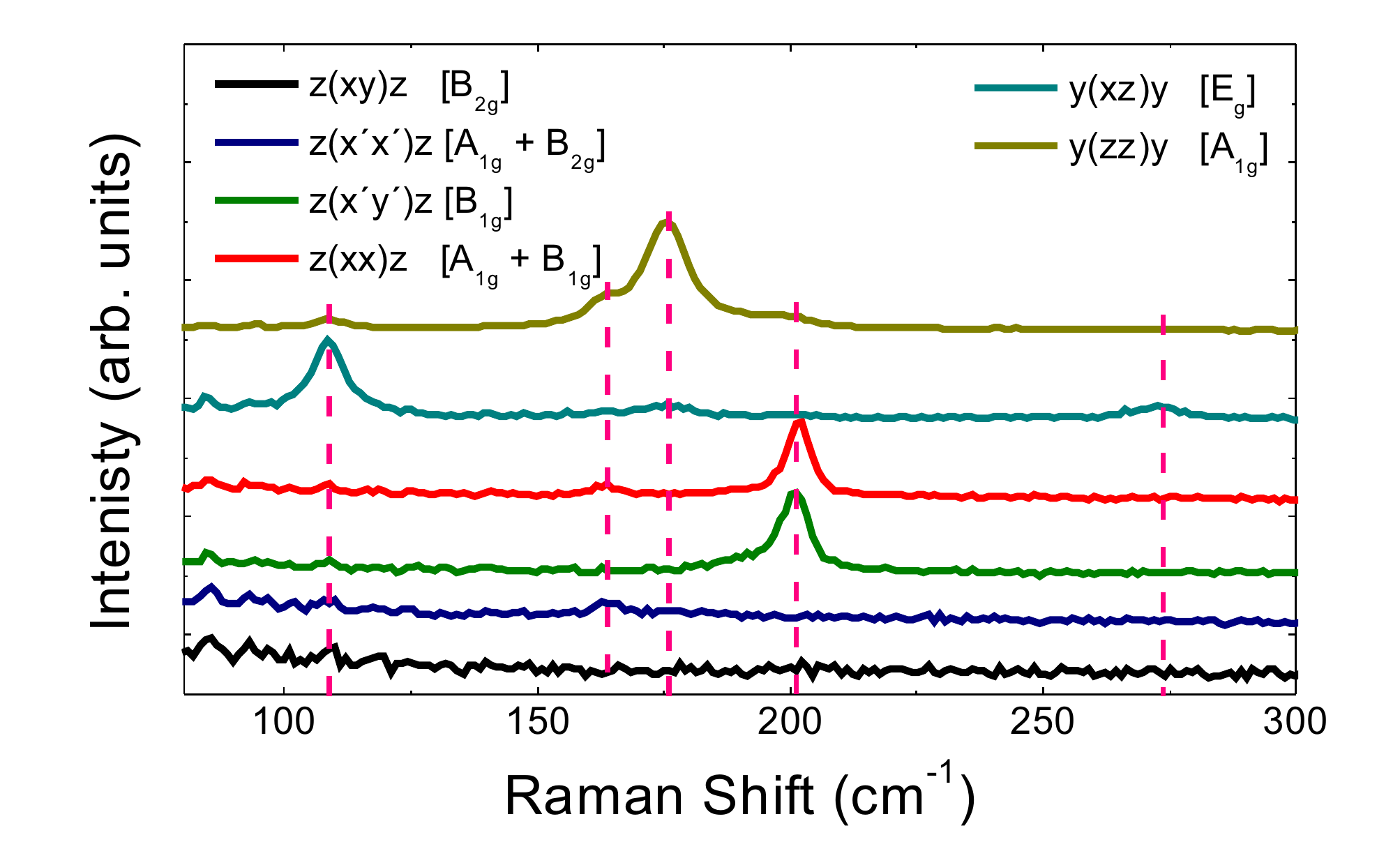}
\caption{Room temperature Raman spectra on NaFeAs in $z(xy)z$, $z(x^{\prime}y^{\prime})z$, $z(x^{\prime}x^{\prime})z$, $z(xx)z$, $y(xz)y$, and $y(zz)y$ configurations. Spectra have been shifted vertically for clarity. Starting from the left, the peaks are assigned to the E$_{g}$(As), A$_{1g}$(As), A$_{1g}$(Na), B$_{1g}$(Fe) and E$_{g}$(Fe) modes.}
\label{Fig1}
\end{figure}

NaFeAs crystallizes in the $P4/nmm$ ($D_{4h}^7$) space group, with the Na, Fe and As atoms located at the 2c, 2b and 2c Wyckoff positions, respectively.
From group symmetry analysis, 15 zone center optical phonons are expected, among which 2A$_{1g}$, 1 B$_{1g}$ and 3E$_g$ modes are Raman active~\cite{Rousseau_JRS1981}.

In Fig.~\ref{Fig1}, we report Raman spectra measured on parent NaFeAs at room temperature for several scattering geometries, with incident light wave vectors along the c-axis [in Porto notations: $z(xy)z$ (that selects excitations with $B_{2g}$ symmetry), $z(x^{\prime}y^{\prime})z$ (B$_{1g}$), $z(x^{\prime}x^{\prime})z$ (A$_{1g}$ + B$_{2g}$), $z(xx)z$ (A$_{1g}$ + B$_{1g}$)] or along the b-axis [$y(zz)y$ (A$_{1g}$) and $y(xz)y$ (E$_{g}$) configurations].

The assignment of the modes is similar to that of LiFeAs reported in ref.~\onlinecite{Um_PRB2012_LFA}. The two modes at 163 cm$^{-1}$ and 213 cm$^{-1}$ in $z(xx)z$ configuration can be assigned to c-axis polarized A$_{1g}$(As) and B$_{1g}$(Fe) vibrations of the FeAs planes. The intense phonon at 178 cm$^{-1}$ in the $y(zz)y$ configuration is considerably suppressed in the $y(xz)y$ configuration, and this phonon can therefore be assigned to the second A$_{1g}$ mode.

These modes frequencies are in good agreement with those calculated within density functional theory in refs.~\onlinecite{Jishi_ACMP2010, Deng_PRB2009}, listed in Table.~\ref{Table_1}.

The two remaining modes at 113 cm$^{-1}$ and 281 cm$^{-1}$ seen experimentally in the $y(xz)y$ configuration are attributed to E$_{g}$ modes.
Comparison with the theoretical estimates allows one to assess the former to the E$_{g}$(As). The energy of the second mode, however, is significantly larger than the ones calculated for the two remaining E$_{g}$(Fe) and E$_{g}$(Na) phonons.
We tentatively assign it to the E$_{g}$(Fe) which has the closest energy. It is therefore likely that the E$_{g}$(Na) is not observed. Further experiments with different incident photon wavelengths might help to observe this mode.

\begin{table*}
\centering
\begin{tabular}{>{\centering}p{2cm}|>{\centering}p{2.5cm}|>{\centering}p{2.5cm}|>{\centering}p{2.4cm}|>{\centering}p{2.4cm}|>{\centering}p{1.2cm}|>{\centering}p{1.2cm}|>{\centering}p{1.2cm}|>{\centering\arraybackslash}p{1.2cm}}
\hline
\hline

 Mode     & Polarization      & Selection Rule  & \multicolumn{2}{c|}{Calc. frequency (cm$^{-1}$)}        & \multicolumn{4}{c}{Exp. fitting parameters (cm$^{-1}$)} \\     \cline{4-9}
(atom)    &                   &                 & from Ref.~\onlinecite{Jishi_ACMP2010} & from Ref. ~\onlinecite{Deng_PRB2009} &  $\omega_0$  &  $C$  &  $\Gamma_0$ & $\Gamma$ \\
\hline
E$_{g}$  (As)  & in-plane & $xz$                              &110 & 126 & 113 & 0.8 &     &       \\
A$_{1g}$ (As)  & c-axis   & $xx$, $x^\prime x^\prime$, $zz$   &176 & 174 & 163 & 0.1 & 1.9 & 1.0   \\
B$_{1g}$ (Fe)  & c-axis   & $xx$, $x^\prime x^\prime$         &218 & 227 & 213 & 2.7 & 0.3 & 0.9   \\
E$_{g}$  (Fe)  & in-plane & $xz$                              &241 & 256 & 281 & 3.8 &     &       \\
E$_{g}$  (Na)  & in-plane & $xz$                              &187 & 203 &     &     &     &       \\
A$_{1g}$ (Na)  & c-axis   & $xx$, $x^\prime x^\prime$, $zz$   &199 & 198 & 178 & 0.9 &     &       \\
\hline
\hline

\end{tabular}
\caption{Calculated Raman-active phonon frequencies and selection rules from Refs.~\onlinecite{Jishi_ACMP2010} and~\onlinecite{Deng_PRB2009} and comparison to our experimental results (see text for the definition of the parameters).}
\label{Table_1}
\end{table*}

\subsection{Temperature \& doping dependence of Phonon modes}

\begin{figure}
\includegraphics[width=1\linewidth]{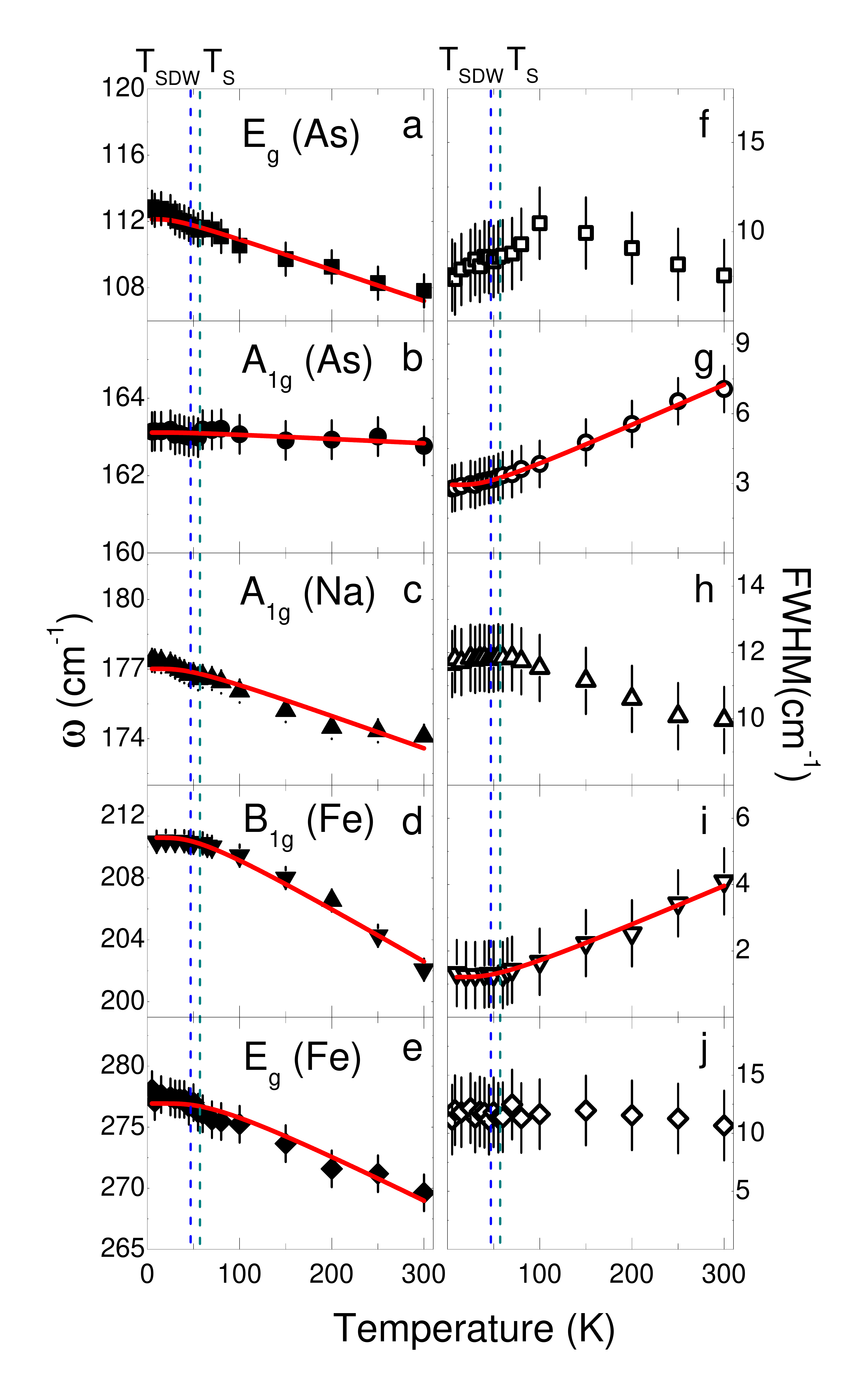}
\caption{a) - e) Temperature dependence of the phonon frequencies, and f) - j) linewidths in NaFeAs. Pink dashed lines correspond to $T_{SDW}$ and $T_{S}$.}
\label{Fig2a}
\end{figure}

\begin{figure}
\includegraphics[width=1\linewidth]{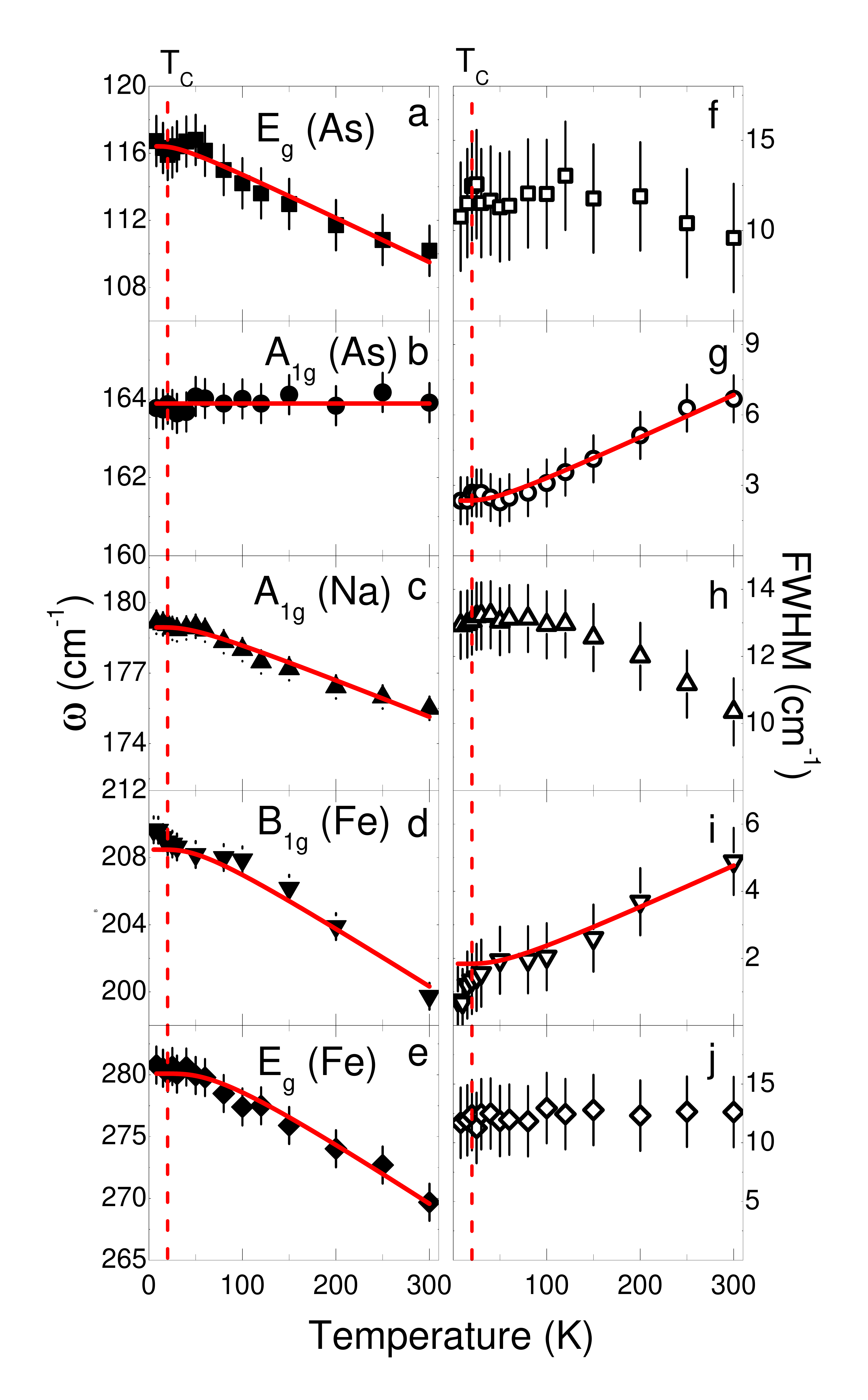}
\caption{a) - e) Temperature dependence of the phonon frequencies, and f) - j) linewidths in NaFe$_{0.97}$Co$_{0.03}$As. Pink dashed line corresponds to $T_c$.}
\label{Fig2b}
\end{figure}

The Raman data have been collected between 5 K and room temperature.
The temperature dependence of the frequencies and linewidths of the five phonons in NaFeAs and NaFe$_{0.97}$Co$_{0.03}$As are reported in
Figs.~\ref{Fig2a} and~\ref{Fig2b}, respectively.
The general trend is a hardening and a narrowing of the modes upon cooling. Such a trend is expected in the absence of effects other than regular anharmonicity.
The temperature dependence of frequencies and linewidths were fitted using a simple symmetric anharmonic phonon decay~\cite{Klemmens_PR1966, Menendez_PRB1984}:

\begin{equation}\label{e1}
\omega_{ph}(\textit{T}) = \omega_{0} - C\bigg[1+\frac{2}{e^{\frac{\hbar\omega_{0}}{2k_{B}T}}-1}\bigg]
\end{equation}
\begin{equation}\label{e2}
\Gamma_{ph}(\textit{T}) = \Gamma_{0} + \Gamma\bigg[1+\frac{2}{e^{\frac{\hbar\omega_{0}}{2k_{B}T}}-1}\bigg]
\end{equation}

\noindent where $C$ and $\Gamma$ are positive constants, $\omega$$_{0}$ is the bare phonon frequency, and $\Gamma$$_{0}$ a residual (temperature independent) linewidth. The obtained fitting parameters are summarized in Table.~\ref{Table_1}.
As shown on Fig.~\ref{Fig2a}, this simple model could only reproduce the temperature dependence of the A$_{1g}$(As) and B$_{1g}$(Fe) of NaFeAs. 
Interestingly, these modes are not renormalized through the structural and magnetic transitions in sharp contrast with the observations made in the 122 compounds, where a clear narrowing of the A$_{1g}$(As) and B$_{1g}$(Fe) phonons has been reported~\cite{Chauviere_PRB2009,Choi_PRB2008,Rahlenbeck_PRB2009, Choi_JPCM2010}. The linewidths of the A$_{1g}$(As) and E$_{g}$(Fe) modes increases upon cooling (Figs.~\ref{Fig2a}-h, and j). This is also the case for the E$_{g}$(As) mode (Figs.~\ref{Fig2a}-f), however below 100 K, this trend is reversed and the linewidth starts to sharpen. 

\begin{figure}
\includegraphics[width=1\linewidth]{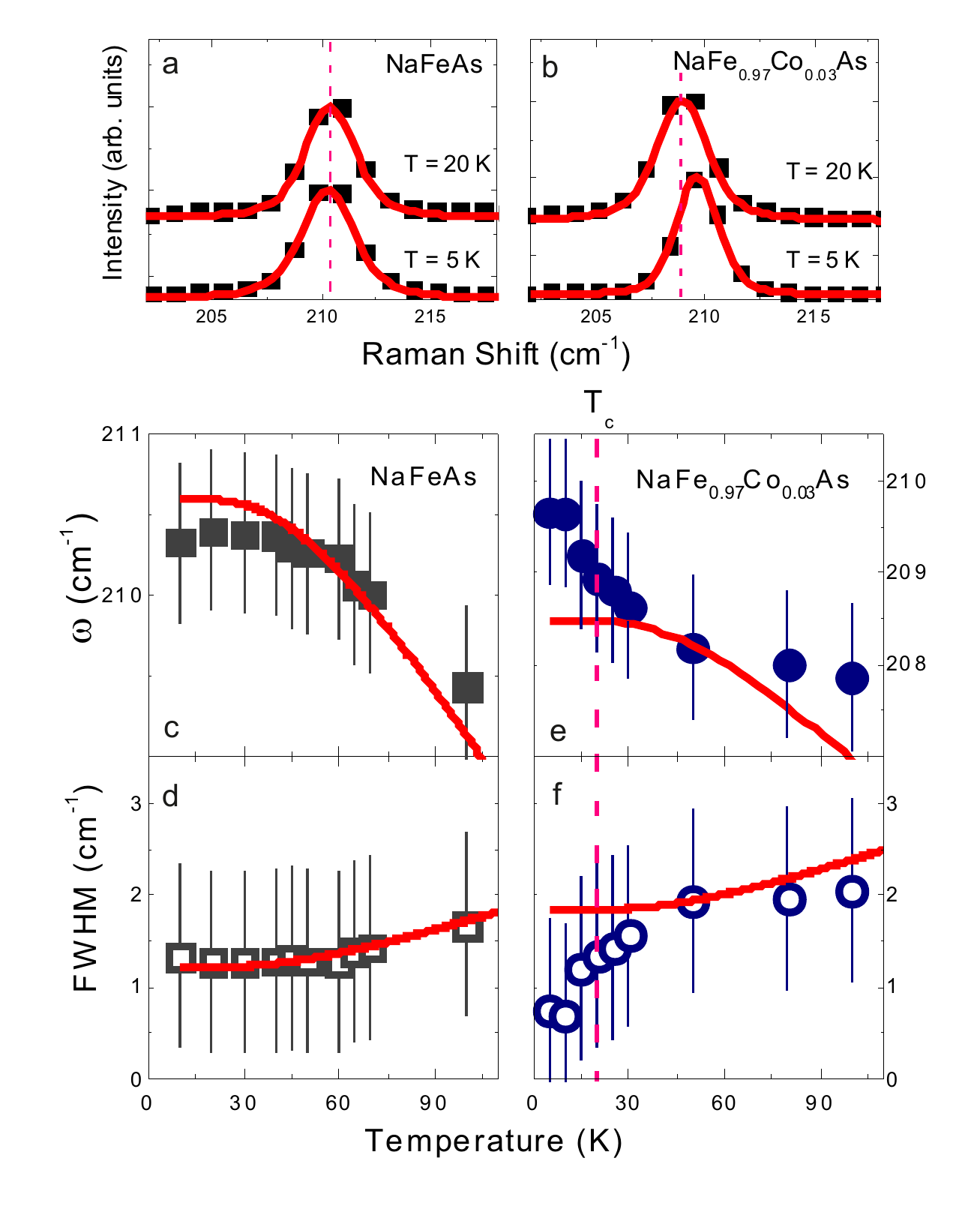}
\caption{ a) Normalized Raman spectra of the B$_{1g}$(Fe) for 5 K and 20 K for NaFeAs. Black squares are the raw data, red line is the fit. The spectra have been shifted vertically for clarity. b) Same plot for NaFe$_{0.97}$Co$_{0.03}$As. c) and d) Low temperature frequencies and linewidths of B$_{1g}$(Fe) mode for NaFeAs. e) and f) Same plot for NaFe$_{0.97}$Co$_{0.03}$As. The pink dashed line marks $T_c$ and the red line is the result of a conventional phonon anharmonic model (see text).}
\label{Fig4}
\end{figure}

In the Co-doped compound, similar results are obtained on the linewidths (Figs.~\ref{Fig2b}-f to j).
In addition, a small, but clear (Fig.~\ref{Fig4}) superconductivity-induced renormalization of the B$_{1g}$(Fe) phonon across the superconducting transition in NaFe$_{0.97}$Co$_{0.03}$As is observed. This renormalization consists in a hardening of about 1.5~cm$^{-1}$ accompanied with a narrowing of similar amplitude of the mode lineshape. This behavior clearly departs from the expected anharmonic behavior extrapolated from the high temperature data.

Beyond this effect on the B$_{1g}$ mode, that will be discussed in details later, the only noticeable differences with the parent compound are i) a significantly broader linewidth of the A$_{1g}$(Na) mode at low temperature, and ii) a weaker narrowing of the E$_{g}$(As) below 100 K.
Finally we note that all the modes, except the B$_{1g}$(Fe), are harder in the doped compound than in the parent one, which can be quantitatively accounted for by the shrinking of the unit cell dimensions upon Co substitution~\cite{Parker_PRL2010}.


\section{Discussion}

\begin{figure*}
\noindent
\includegraphics[width=1\linewidth]{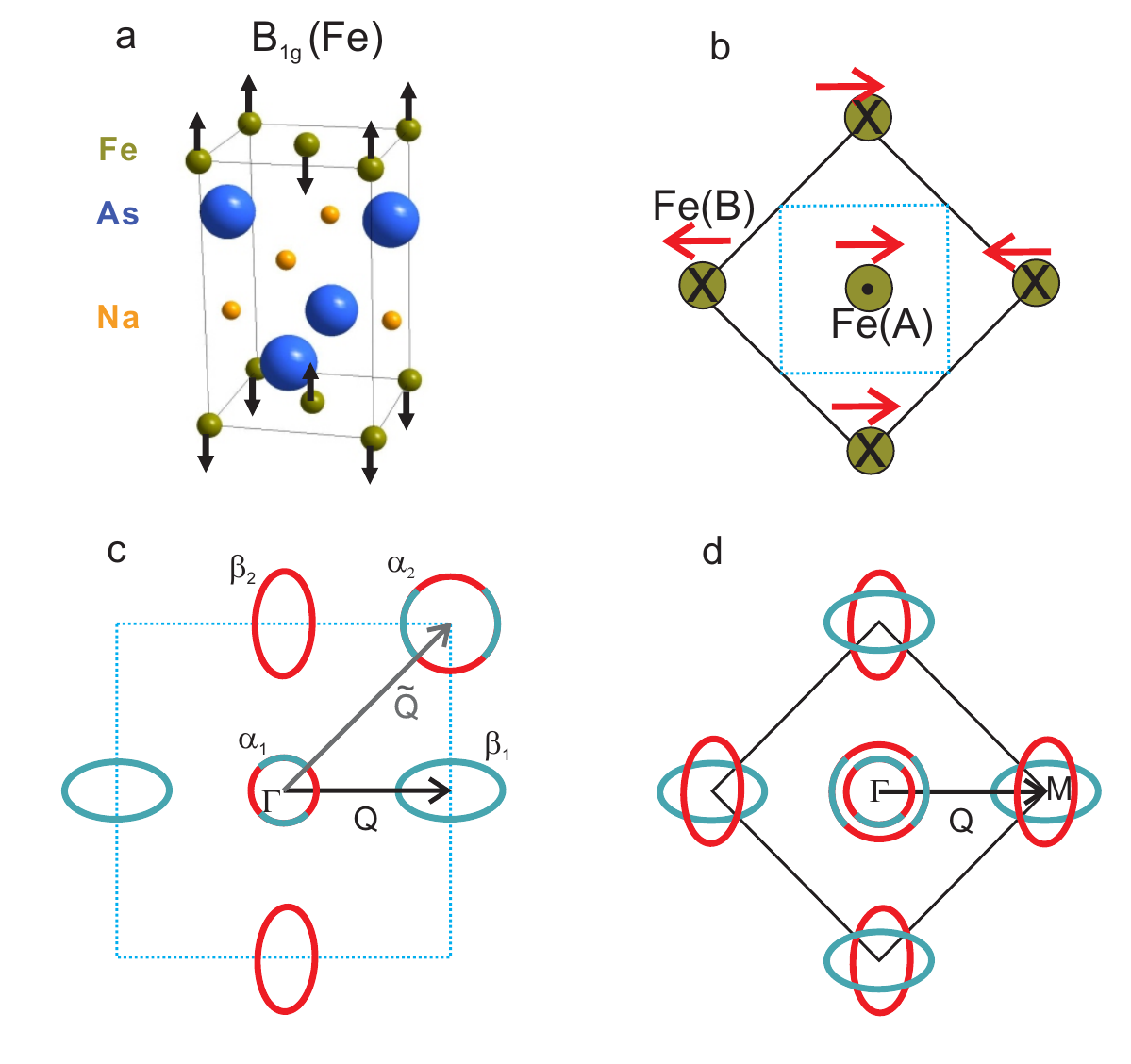}
\caption{(Color online) a) Pattern of the B$_{1g}$(Fe) phonon mode, b) planar projection of the Fe plane as well as the SDW order (red arrows).
Dashed line of the smaller square is the one-Fe unit cell.
c) The corresponding Brillouin zones of the one- and d) two-Fe unit cells and the SDW wave vector ${\bf Q}$ connecting the hole band (circular shapes) and the electron band (ellipsoidal). The color code indicates the orbital character (green for $d_{zx}$ and red for $d_{zy}$) of the bands in a minimal two bands model~\cite{Hirschfeld_RPP2011}. The vector ${\bf \vec{Q}}=(\pi,\pi)$ in panel c corresponds to the unit cell doubling that must be taken into account when considering the $B_{1g}$ vibration, while ${\bf Q}$ represents the SDW wave vector, that connects nested portions of the electron and hole pockets.
\label{fig5}}
\end{figure*}

In previous phonon studies on iron pnictides, the narrowing of the phonon linewidth has been observed through the SDW transition~\cite{Um_PRB2012_FTS,Gnezdilov_PRB2011,Chauviere_PRB2009,Choi_PRB2008,Rahlenbeck_PRB2009, Choi_JPCM2010}. In those cases, the B$_{1g}$(Fe) mode (that corresponds to the c-axis polarized vibration of planar Fe (Fig.~\ref{fig5}-a)) and the A$_{1g}$(As) (or A$_{1g}$(Te) in the 11 compounds) modes are generally the most affected. This is understood as a consequence of a suppression of the electron-phonon coupling as the electronic density of states at the Fermi level decreases when the SDW gap opens. Splitting of the in-plane, doubly degenerate, E$_g$ vibrations is also observed at the structural transition~\cite{Chauviere_PRB2009}. The large amplitude of the latter effect is associated with strong magneto-elastic effects.
Despite the similarity of the structural/magnetic phase transitions undergone by NaFeAs and 122 compounds at low temperature, none of these effects are seen here.

Instead, the B$_{1g}$(Fe) and A$_{1g}$(As) modes behave exactly like in LiFeAs~\cite{Um_PRB2012_LFA}, and display a continuous hardening and sharpening down to the lowest temperatures. The small residual linewidths $\Gamma_0$ of these modes indicate good sample homogeneity.
This contrasts with the behavior of the three other modes, as shown in Fig.~\ref{Fig2a}-f, h and j. The most interesting behavior is arguably the one of the E$_{g}$(As) mode at 110 cm$^{-1}$ that broadens upon cooling and then sharpens below $\sim$ 100 K, a temperature larger than both T$_{SDW}$ (closure of the SDW gap at 40 K has recently been reported in the parent compound~\cite{Zhou_PRL2012}) and T$_{S}$~\cite{He_PRL2010}. We note that recently, electron nematicity above these transition temperatures has recently been reported in NaFeAs~\cite{Rosenthal_NatPhys2014}. Further investigations are required to check whether these two phenomena are actually related.

The behavior of the A$_{1g}$(Na) and of the E$_{g}$(Fe) modes, that broaden down to base temperature, is reminiscent of the one of the A$_{1g}$(Te) polarized modes in parent Fe$_{1+y}$Te compound~\cite{Gnezdilov_PRB2011,Um_PRB2012_FTS}. In these compounds, it has been argued that it originates from anharmonicity associated with the combined effect of spin-orbital frustration and large Fe magnetic moments~\cite{Gnezdilov_PRB2011}. This doesn't seem to be the case here since the magnetic moment in the ordered phase is rather modest ($\sim$ 0.1 $\mu_B$~\cite{Li_PRB2009, Wright_PRB2012}). Furthermore, the A$_{1g}$(As) mode, that is in principle more sensitive to this effect since it modulates the Fe-As distance that controls magnetic order~\cite{Yildirim_PhysicaC2009}, is not affected. The effect is also clearly present in both parent and doped compounds. This implies that it might been intrinsically related to the anharmonic potential experienced by Na atoms between the two FeAs layers rather than to the details of the electronic structure of these layers.

This is not the case for the B$_{1g}$(Fe) phonon which undergoes a doping-dependent, superconductivity-induced renormalization that we will discuss in details in the following section.

\section{Theoretical Modeling for the B$_{1g}$ mode superconductivity-induced renormalization}

Superconductivity induced phonon renormalizations are in principle directly related with the redistribution of the electronic states, and therefore of the electron-phonon coupling, as the superconducting gap opens. In most of the Fe-based superconductors, no specific changes were reported in the phonon frequencies and linewidths across $T_c$ ~\cite{Litvinchuk_PRB2008,Gallais_PRB2008,Rahlenbeck_PRB2009,Um_PRB2012_LFA}, which has been attributed to the small superconducting gap amplitude 2$\Delta$ compared to the phonon frequencies.
Interestingly, similar renormalizations (\textit{i.e.} hardening + narrowing of comparable amplitudes) of the $B_{1g}$ phonon across $T_c$ have been reported in Ba$_{1-x}$K$_x$Fe$_2$As$_2$, Sr$_{1-x}$K$_x$Fe$_2$As$_2$~\cite{Choi_JPCM2010} or Pr$_x$Ca$_{1-x}$Fe$_2$As$_2$ ~\cite{Litvinchuk_PRB2011}.
As pointed out by the authors of ref.~\onlinecite{Choi_JPCM2010}, in the conventional approach of the electron-phonon coupling, even considering the s$_{\pm}$-wave symmetry of the superconducting gap, the narrowing of the phonon is puzzling since it would imply that 2$\Delta$ is unrealistically larger than the energy of the renormalized phonon energy (210 cm$^{-1}$ $\sim$ 26.4 meV). Recent estimations of 2$\Delta$ in optimally doped NaFe$_{0.97}$Co$_{0.03}$As range from 2$\Delta$ = 10.72 meV ($\sim$ 86.5 cm$^{-1}$) from specific heat~\cite{Wang_PRB2012}, 2$\Delta$ = 11 meV ($\sim$ 88 cm$^{-1}$) from STM~\cite{Zhou_PRL2012}, and 2$\Delta$ = 13.6 meV ($\sim$ 109.7 cm$^{-1}$) from ARPES~\cite{Liu_PRB2011} and indeed confirm that the mode renormalization cannot be understood within the conventional framework.
Choi et al.~\cite{Choi_JPCM2010} also suggested that the two missing ingredients might be the effects of interband scattering and of the coexistence of superconductivity with SDW.

The physical idea behind the model calculation presented hereafter, is that interband scattering at ${\bf Q}$=0 connects only the matrix element between orthogonal orbital states, leading to vanishing transition probability. The inclusion of the SDW order (or say long range correlations) opens new channels for ${\bf Q}$=($\pi$,0) inter-band scattering, which connects the non-orthogonal orbital states, and therefore leads to finite transition probability, that renormalizes the phonon through the superconducting transition (in addition, a sign-changing ($s_{+-}$) superconducting order parameter between hole and electron-pockets is required to get the correct signs in the renormalization).

In the case of NaFe$_{0.97}$Co$_{0.03}$As, no signature of static SDW have been reported, yet from the proximity with the SDW phase boundary, one can naturally expect the presence of sizeable SDW correlations that might play similar role at any frequency larger than the spin gap $\Omega_{SDW}$.
Inelastic Neutron scattering measurements performed in compounds with close chemical composition~\cite{Zhang_PRL2013} show that this gap in the spin excitation spectra is a couple of meV large, indicative of rather slow SDW fluctuations. Since this energy scale is in any event much smaller than the frequency of the B$_{1g}$ phonon we are interested in, we will treat the SDW order as static in the following.

\subsection{Model details}
First we notice that the Fe-Fe plane has no mirror symmetry above
and below the plane. In analogy with the CuO$_2$ planes in superconducting YBa$_2$Cu$_3$O$_{6+x}$~\cite{Devereaux1995}, this asymmetric environment results in a local electric field along the c-axis, ${\bf E^z}$.
Then, the alternating displacements of the Fe atoms along the c-axis due to the B$_{1g}$
vibration (Fig.~\ref{fig5}-a) lead to a spatial modulation of the energy of the $d$-electrons which form the conduction bands and also the SDW order $\Delta_{SDW}$.
Following ref.~\cite{Devereaux1995}, we write the electron-phonon interaction for the B$_{1g}$ phonon as
\begin{eqnarray}
H_{ph-elec} = e \sum_{n\in A, m \in B; \tau} [{\bf E_A ^z u_A}(a{\bf n}) d^{\dag}_{n,\tau} d_{n,\tau} \\ \nonumber
+ {\bf E_B ^z u_B}(a{\bf m}) d^{\dag}_{m,\tau} d_{m,\tau}]
\label{eq:epc}
\end{eqnarray}
\noindent where ${\bf A}$ and ${\bf B}$ stand for the alternating A- and
B-sites of Fe-atoms (see Fig.~\ref{fig5}-b), and $\tau$ stands for both the spin and
orbital degrees of freedom. ${\bf b_A}(a{\bf n})$ and ${\bf
b_B}(a{\bf m})$ indicate the displacement vectors of Fe(A) and
Fe(B), respectively. For the B$_{1g}$ phonon, we assume ${\bf E_A ^z =
E_B ^z}$ and ${\bf u_A} = - {\bf u_B}$.
The energy shift of the $d$-orbitals couples to the charge density as well as to the amplitude of the SDW order $\Delta_{SDW}$.
Therefore, in general we can assume two possible forms of the electron-phonon coupling in the compound as:
\begin{equation}
H_{ph-den} =  \sum_{\bf q}  g_{\bf q} (b_{\bf q} + b^{\dagger}_{-{\bf q}} ) \sum_{\bf k} \Psi^{\dagger}_{\bf k} \mathbb{1} \Psi_{{\bf k+q}}
\label{eq:phden}
\end{equation}
\begin{equation}
H_{ph-SDW} = \sum_{\bf q} g_{\bf q} (b_{\bf q} + b^{\dagger}_{-{\bf q}} ) \sum_{\bf k} \Psi^{\dagger}_{\bf k} \tau_1 \Psi_{{\bf k+q}}
\label{eq:phsdw}
\end{equation}
\noindent where $b_{\bf q}$ and $b^{\dagger}_{-{\bf q}}$ are the phonon annihilation and creation operators and  $\Psi_k = (c_{\bf k}, c_{\bf k+Q})$ with ${\bf Q}=(\pi,0)$ is the general spinor
representation for the SDW order. $c_{\bf k}$ is the short hand notation of the annihilation operator of the band electrons for all bands
defined in folded BZ (the two-Fe cell BZ depicted in Fig.~\ref{fig5}-d) and the spin index is omitted because the above form just duplicates for both spins.

Eq.~\ref{eq:phden} above describes the ordinary coupling of a phonon to the charge density.
We, however, will argue below that this coupling term is very weak as far as the B$_{1g}$ phonon is concerned.
The central idea is that the alternating displacement of the neighboring Fe-atoms in the B$_{1g}$ vibration depicted in Fig.~\ref{fig5}-a and Fig.~\ref{fig5}-b,
requires to work in the folded BZ (Fig.~\ref{fig5}-d) rather than in the non-folded one (Fig.~\ref{fig5}-c).
There, the ${\bf q=0}$ density operator $\Psi^{\dagger}_{\bf k} \mathbb{1} \Psi_{{\bf k}}$  
connects a large momentum scattering process with the exchange momenta ${\bf \vec{Q}}=(\pi,\pi)$ in the one-Fe unit cell BZ (distinct from the SDW ordering ${\bf Q}$ vector, see  Fig.~\ref{fig5}-c).

This large ${\bf \vec{Q}}$ vector becomes ${\bf \vec{Q}}=(0,0)$ after the BZ folding for the two-Fe unit cell, hence the B$_{1g}$ phonon looks like coupling
with an ordinary charge density of a small momenta modulation $\Psi^{\dagger}_{\bf k} \mathbb{1} \Psi_{{\bf k}}$.
However, in more microscopic model analysis, it is necessary to introduce at least two orbitals (full description needs five d-orbitals)
degrees of freedom per Fe-site ($d_{zx}$ and $d_{zy}$ orbitals)~\cite{Hirschfeld_RPP2011,Raghu_PRB2008, Graser_NJP2009} to properly describe this folding process
which backfolds ${\bf \vec{Q}}=(\pi,\pi)$ of the one-Fe unit cell BZ into zone center $\Gamma$ of the two-Fe unit cell BZ.

In the minimal two orbital model, before folding there exist one hole band
($\alpha_1$ band) around $\Gamma=(0,0)$ and another hole band ($\alpha_2$ band) around $\tilde{\Gamma}=(\pi,\pi)$, and one
electron band ($\beta_1$ band) around $(\pi,0)$ and one electron band ($\beta_2$ band) around $(0,\pi)$ (Fig.~\ref{fig5}-c).

The scattering process described above by the standard charge density coupling
$\Psi^{\dagger}_{\bf k} \mathbb{1} \Psi_{{\bf k}}$ actually corresponds to interband scattering processes of $\alpha_1 \Leftrightarrow \alpha_2$ and $\beta_1 \Leftrightarrow \beta_2$.
Then it is easy to notice from the orbital contents ($d_{zx}$ and $d_{zy}$) of the bands $\alpha_{1,2}$, and $\beta_{1,2}$, that the scattering processes $\alpha_1 \Leftrightarrow \alpha_2$ and $\beta_1 \Leftrightarrow \beta_2$ are dominated by the
inter-orbital scattering in terms of their orbital contents, namely $d_{zx} \Leftrightarrow d_{zy}$. This coupling term to B$_{1g}$ phonon vanishes or becomes very weak, and therefore we will not consider the term $\Psi^{\dagger}_{\bf k} \mathbb{1} \Psi_{{\bf k}}$ any more.

On the other hand, Eq.~\ref{eq:phsdw} with $\Psi^{\dagger}_{\bf k} \tau_{1} \Psi_{{\bf k}}$ connects the
scattering processes of $\alpha_{1,2} \Leftrightarrow \beta_{1,2}$ with the help of the SDW correlation.
Notice that the scattering processes of $\alpha_{1,2} \Leftrightarrow \beta_{1,2}$ is the scattering between
the hole band and the electron bands and vice-versa. Also it is easy to find from the orbital content of the
bands that these processes contain a large amplitude of the intra-orbital scattering processes such as $d_{zx} \Leftrightarrow
d_{zx}$ and $d_{zy} \Leftrightarrow d_{zy}$. The electron-phonon interaction Hamiltonian (Eq.~\ref{eq:phden}) shows that only intra-orbital
scattering processes will couple with the B$_{1g}$ phonon.
Therefore, from now on we will consider the $\Psi^{\dagger}_{\bf k} \tau_{1} \Psi_{{\bf k}}$ coupling term of Eq.~\ref{eq:phsdw} only.

Rewriting Eq.~\ref{eq:phsdw} in the minimal two band model
of Fe-pnictide compounds, for example, for the model band shown in
Fig.~\ref{fig5}-c, we obtain:
\begin{equation}
H_{ph-SDW} = \sum_{\bf q} g_{\bf q} (b_{\bf q} + b^{\dagger}_{-{\bf q}} ) \sum_{\bf k} \Big[ h^{\dagger}_{\bf k}  e_{{\bf k+Q+q}} + h.c. \Big]
\end{equation}
\noindent where $h_k$ describes a hole band around $\Gamma$ point and $e_k$ describes an electron band around $M$ point (Fig.~\ref{fig5}-d). Using the
above coupling $H_{ph-SDW}$, it is straightforward to calculate the phonon self-energy and examine the change of the phonon
frequency and its linewidth due to the change of the electronic excitations.  Neglecting vertex corrections, the renormalization
of optical phonons is determined by conventional Dyson equation:
\begin{equation}
D^{-1}({\bf q},\omega) = D_0 ^{-1}({\bf q},\omega) + g_{{\bf q}}^2 \Pi({\bf q},\omega)
\label{eq:dyson}\end{equation}
\noindent with the bare phonon propagator $D_0 ^{-1}(\omega)=\frac{\omega^2-\omega_0 ^2 + i \delta}{2 \omega_0}$, the
electron-phonon coupling $g_{{\bf q}}$ of Eq.~\ref{eq:dyson} and the relevant polarizability $\Pi({\bf q},\omega)$ given by:
\begin{eqnarray}
\Pi_{he}({\bf q},\omega) &=& i \int Tr \Big[ \hat{\tau}_3  \hat{G}_h({\bf k}, \Omega+\omega) \times  \nonumber \\
& & \hat{\tau}_3  \hat{G}_e({\bf k+q+Q}, \Omega) \Big] \frac{d^2k d^2\Omega}{(2 \pi)^3}.
\label{eq:Pieh}
\end{eqnarray}

where the Green's functions of the electron band and hole band are:
\begin{equation}
\hat{G}_{b = h, e} ({\bf k}, \Omega) = \frac{ \omega \hat{\tau}_0 +
\epsilon_{b}({\bf k}) \hat{\tau}_3 + \Delta_b({\bf k}) \hat{\tau}_1 }
{\omega^2 - \epsilon_{b}^2({\bf k})- \Delta_b^2({\bf k}) + i \delta}
\end{equation}
\noindent $\tau_i$ are the Pauli matrices, $\epsilon_{b}({\bf k})$ is the band dependent electronic dispersion and $\Delta_b({\bf k})$ is the superconducting gap, that vanishes in the normal state. In our calculation, we assume that the compound has the typical sign-changing s-wave pairing state ($s_{\pm}$) and
therefore each band has a constant s-wave gap but with opposite signs to each others as $\Delta_h > 0 $ and $\Delta_e < 0$.

The renormalization of the Raman phonon frequency (resp. linewidth) is obtained by calculating the difference between the real (resp.
imaginary) part of the electronic polarizability in the superconducting and normal states: $\Delta\Pi ({\bf q},\omega) =\Pi^S ({\bf q},\omega) - \Pi^N
({\bf q},\omega)$ for ${\bf q = 0}$.

The ordinary intraband polarizability $\Pi^S_{bb}$ (where $b = h$ or $e$ depending on the chosen band) in the superconducting state has the following standard coherence
factor for the pair scattering process:
\begin{equation}
\Pi^S_{bb} ({\bf q}, \Omega) \approx \Big(1 - \frac{\epsilon_b({\bf k}) \epsilon_b({\bf k+q})
- \Delta_b({\bf k}) \Delta_b({\bf k+q}) }{E_{{\bf k}}E_{{\bf k+q}}}\Big).
\end{equation}

\noindent For interband scattering on the other hand, $\Pi^S_{h,e}$, since in our model the gap functions on each band have
opposite sign, the coherence factor for the polarizability is:
\begin{eqnarray}
\Pi^S_{he}  ({\bf q}, \Omega) & \approx & \Big(1 - \frac{\epsilon_h({\bf k}) \epsilon_e({\bf k+q+Q})
- \Delta_h \Delta_e }{E_{{\bf k}}E_{{\bf k+q+Q}}}\Big) \\
 & =& \Big(1 - \frac{\epsilon_h({\bf k}) \epsilon_e({\bf k+q+Q})
+  |\Delta_h| |\Delta_e | }{E_{{\bf k}}E_{{\bf k+q+Q}}}\Big)
\end{eqnarray}

\subsection{Results}

\begin{figure*}
\noindent
\includegraphics[width=1\linewidth]{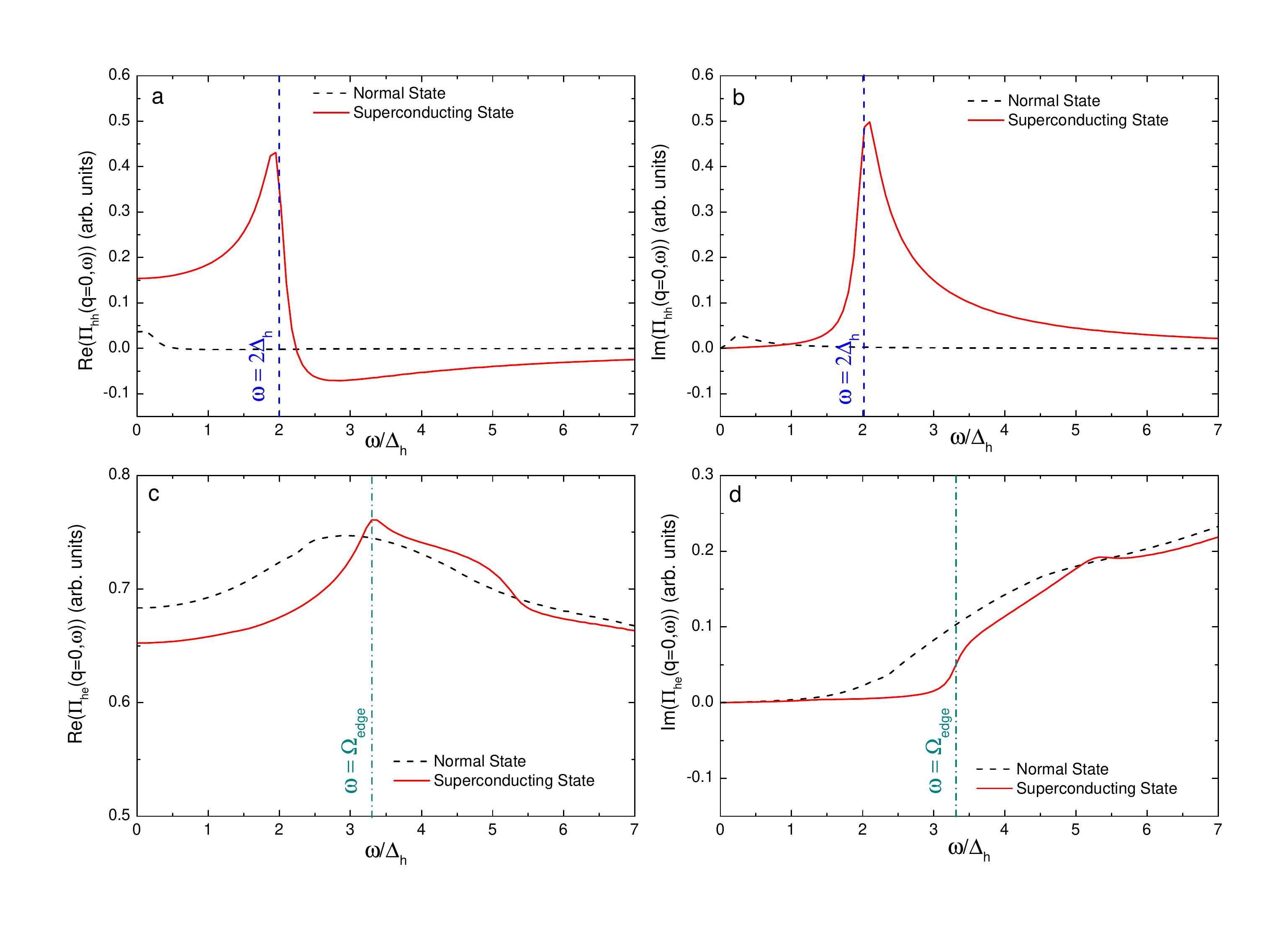}
\vspace{0cm}
\caption{(Color online) Real and Imaginary parts of polarizability $\Pi({\bf q=0},\omega)$
in Normal and Superconducting state. (a-b) Standard intraband polarizability $\Pi_{(hh)}$.
(c-d) Interband polarizability $\Pi_{(he)}$.
\label{fig6}}
\end{figure*}

On Fig.~\ref{fig6}, we show the results of the calculation at {\bf q=0} for the real and imaginary parts of $\Pi_{hh}$ and $\Pi_{he}$ as a function
of the Raman shift $\omega$, in the normal and superconducting states.

The behavior of the phonon across the superconducting transition taking into account intra-band scattering $\Pi_{hh}$ only follows the conventional trend for a s-wave superconductor~\cite{Zeyher_ZPB1990} that is:\\
(1) For $\omega > 2 \Delta_h$, the real (resp. imaginary) part of $\Delta\Pi_{hh} ({\bf q = 0},\omega)$ is negative (resp. positive), corresponding to a hardening (resp. broadening)
 of the phonon frequency (resp. linewidth). \\
(2) For $\omega <<  2 \Delta_h$, the real (resp. imaginary) part of $\Delta\Pi_{hh} ({\bf q = 0},\omega)$ is positive (resp. negative), corresponding to a softening (resp. sharpening) of the phonon frequency (resp. linewidth).\\

As pointed out earlier, this behavior contrasts with the observation of a hardening (resp. sharpening) of the B$_{1g}$ phonon frequency (resp. linewidth), despite the fact that we clearly have $\omega_{B1g} > 2 \Delta_h$.

However, since the B$_{1g}$ phonon couples to the electrons only through Eq.~\ref{eq:dyson}, the relevant polarizability is the interband $\Pi$= $\Pi_{he}$ one rather than $\Pi_{hh}$.
The results of the calculation for the real and imaginary parts of $\Pi_{he}$ are shown in Figs.~\ref{fig6}-c and -d.
There is a wide region of frequencies where the real part of $(\Delta\Pi_{he} ({\bf q = 0},\omega)$ is negative, which corresponds to a  hardening of the phonon frequency across the superconducting transition. Furthermore, in this region, the imaginary part of $\Delta\Pi_{he} ({\bf q = 0},\omega)$ is also negative indicating a phonon sharpening.

The energy threshold below which this behavior occurs is not set anymore by the superconducting gap amplitudes, but by $\Omega_{edge} = |\Delta_h| + |\Delta_e| + \Delta \epsilon$, where $\Delta \epsilon = min |\epsilon_h ({\bf k})-\epsilon_e ({\bf k +Q})|$.
This quantity vanishes only in the case of perfect nesting between the electron-band and the hole-one shifted by {\bf Q}. Since generally this nesting is not perfect, $\Delta \epsilon$ is non-zero, and when large enough it allows $\Omega_{edge} > \omega_{B1g}$. In this case, a hardening and narrowing of the B$_{1g}$ phonon such as the one experimentally observed in the Co-doped crystal, can be simultaneously obtained across the superconducting transition.
We finally recall that the necessary conditions to obtain these results are both the SDW correlations (to allow non-vanishing interband scattering) and the sign-change of the superconducting gap between hole and electron-pockets.

\section{Conclusion}

In summary, a detailed Raman scattering study has been carried out on NaFe$_{1-x}$Co$_{x}$As (x = 0, 0.03). The modes have been assigned and their temperature dependence studied. In the Co-doped compound, a non-standard renormalization of the Fe B$_{1g}$(Fe) mode across the superconducting transition has been observed. This renormalization can not be understood within a single band and simple multiband approaches. We show that the additional inclusion of sufficiently strong SDW correlations, along with a sign-changing s-wave pairing state, can lead to the experimentally observed behavior.

\section*{Acknowledgements}

We acknowledge  Armin Schulz for technical help, Bernhard Keimer for support and reading of the manuscript, Hlynur Gretarsson and Michaela Souliou for critical reading of the paper, and the Leading Foreign Research Institute Recruitment Program (Grant No. 2012K1A4A3053565) through the National Research Foundation of Korea (NRF) funded by the Ministry of Education, Science and Technology (MEST).
YB was supported by Grants No. NRF-2011-0017079 and No.2013-R1A1A2-057535 funded by the NRF of Korea.
YSK was supported by the Basic Science Research Program (NRF-2013R1A1A2009778) and the Leading Foreign Research Institute Recruitment Program (Grant No. 2012K1A4A3053565).


\end{document}